\documentclass[pra,superscriptaddress,showpacs,twocolumn,10pt]{revtex4}
\usepackage{amssymb}
\usepackage{graphicx}

\begin{document}%

\newcommand{\ket}[1]{|#1\rangle}
\newcommand{\bra}[1]{\langle#1|}
\newcommand{\lr}\longrightarrow
\newcommand{\ra}\rightarrow
\newcommand{\tr}{{\rm Tr}}
\newcommand{\fsp}{{\rm supp}}

\newtheorem{Theorem}{Theorem}
\newtheorem{Prop}{Proposition}
\newtheorem{Coro}{Corollary}
\newtheorem{Lemma}{Lemma}
\title{Unambiguous discrimination between quantum mixed states}
\author{Yuan Feng}
\email{fengy99g@mails.tsinghua.edu.cn}
\author{Runyao Duan}
\email{dry02@mails.tsinghua.edu.cn}
\author{Mingsheng Ying}
\email{yingmsh@tsinghua.edu.cn} \affiliation{State Key Laboratory
of Intelligent Technology and Systems, Department of Computer
Science and Technology Tsinghua University, Beijing, China,
100084}
\date{\today}\begin{abstract}
We prove that the states secretly chosen from a mixed state set
can be perfectly discriminated if and only if these states are
orthogonal. The sufficient and necessary condition when
nonorthogonal quantum mixed states can be unambiguously
discriminated is also presented. Furthermore, we derive a series
of lower bounds on the inconclusive probability of unambiguous
discrimination of states from a mixed state set with \textit{a
prior} probabilities.
\end{abstract}
\maketitle

Quantum state discrimination is an essential problem in quantum
information theory. Perfect discrimination among nonorthogonal
pure states is, however, forbidden by the laws of quantum
mechanics. Nonetheless, if a non-zero probability of inconclusive
answer is allowed, one can distinguish with certainty linearly
independent pure states. This strategy is usually called
\textit{unambiguous discrimination}. Unambiguous discrimination
among two equally probable nonorthogonal quantum pure states was
originally addressed by Ivanovic \cite{IV87}, and then Dieks
\cite{DI88} and Peres \cite{PE88}. Jaeger and Shimony \cite{JS95}
extended their result to the case of two nonorthogonal pure states
with unequal priori probabilities. Chefles \cite{CH98} showed that
$n$ quantum pure states can be unambiguously discriminated if and
only if they are linearly independent. For the general case of
unambiguous discrimination between $n$ pure states with \textit{a
prior} probabilities, it was shown in \cite{SZ02} and \cite{EL03}
that the problem of optimal discrimination, in the sense that the
success probability is maximized, or equivalently, the
inconclusive probability is minimized, can be reduced to a
semidefinite programming (SDP) problem, which has only numerical
solution in mathematics. On the other hand, Zhang \textit{et al}
\cite{ZF01} and Feng \textit{et al} \cite{FZ02} derived two lower
bounds on the inconclusive probability of unambiguous
discrimination among $n$ pure states.

Somewhat surprisingly, it is only recently that the problem of
unambiguous discrimination between mixed states is considered. In
Ref. \cite{SB02}, the optimal unambiguous discrimination between a
pure state and a mixed state with rank 2 was examined. Rudolph
\textit{et al} \cite{RS03} derived a lower bound and an upper
bound on the maximal probability of successful discrimination of
two mixed states. Raynal \textit{et al} \cite{RL03} presented two
reduction theorems to reduce the optimal unambiguous
discrimination of two mixed states to that of other two mixed
states which have the same rank. In the general case of $n$ mixed
state discrimination, Fiurasek and Jezek \cite{FJ03} and Eldar
\cite{EL031} gave some sufficient and necessary conditions on the
optimal unambiguous discrimination and some numerical methods were
discussed.

In this paper, we consider first the distinguishability of any
mixed state set. We prove that any state chosen from a mixed state
set can be perfectly discriminated if and only if the set are
orthogonal, in the sense that any state in the set has support
orthogonal to those of the others. For the case of nonorthogonal
mixed state set, the sufficient and necessary condition of when
states from it can be unambiguously discriminated is that any
state in the set has the support space not totally included in the
supports of the others. Furthermore, we consider the problem of
discriminating unambiguously of $n$ mixed states with \textit{a
prior} probabilities and present a series lower bounds on the
inconclusive probability.

Suppose a quantum system is prepared in a state secretly drawn
from a known set $\rho_1,\dots,\rho_n$, where each $\rho_i$ is a
mixed state in the Hilbert space $\mathcal{H}$. The task of
discrimination is to obtain as much information about the
identification of the state as possible. In what follows, by
perfect discrimination we mean that one can always get the correct
answer while by unambiguous discrimination we mean that except a
maybe nonzero inconclusive probability, one can identify the state
without error. It is obvious that perfect discrimination is
necessarily an unambiguous one, but the reverse is not true in
general. To unambiguously discriminate $\rho_1,\dots,\rho_n$, one
can construct a most general positive-operator valued measurement
(POVM) comprising $n+1$ elements $\Pi_0,\Pi_1,\dots,\Pi_n$ such
that
$$\Pi_i\geq 0, \ \ \ i= 0,1,\dots,n$$
\begin{equation}\label{mesure}
\sum\limits_{i=0}^n \Pi_i =I
\end{equation}
where $I$ denotes the identity matrix in $\mathcal{H}$. Each POVM
element $\Pi_i$, $i=1,\dots,n$ corresponds to identification of
the corresponding state $\rho_i$, while $\Pi_0$ corresponds to the
inconclusive answer. For the sake of simplicity, we often specify
only $\Pi_1,\dots,\Pi_n$ for a POVM since the left element $\Pi_0$
is uniquely determined by $\Pi_0=I-\sum_{i=1}^n \Pi_i$. It is then
straightforward that a POVM $\Pi_1,\dots,\Pi_n$, $\sum_{i=1}^n
\Pi_i\leq I$, can perfectly discriminate $\rho_1,\dots,\rho_n$ if
and only if
$$\tr(\rho_i \Pi_j)=\delta_{ij}$$
while can unambiguously discriminate $\rho_1,\dots,\rho_n$ if and
only if
$$\tr(\rho_i \Pi_j)=p_i\delta_{ij}$$
for some $p_i> 0$, where $i,j=1,\dots,n$.

Since the intersection of the kernels of all $\rho_i$,
$i=1,\dots,n$, is not useful for the purpose of unambiguous
discrimination, sometimes we can assume without loss of generality
that each $\Pi_i$, $i=1,\dots,n$, is in
$\fsp(\rho_1,\dots,\rho_n)$. Here $\fsp(\rho_1,\dots,\rho_n)$ is
defined by the Hilbert space spanned by eigenvectors of the
matrices $\rho_1,\dots,\rho_n$ with nonzero corresponding
eigenvalues.

The following lemma is a necessary condition of a POVM to
unambiguously discriminate a given mixed state set.

\begin{Lemma}\label{lemma:1}
Suppose $\Pi_1,\dots,\Pi_n$ are POVM elements and $\sum_i
\Pi_i\leq I$. If for any $i$, $\Pi_i$ can unambiguously
discriminate $\rho_i$, then $\Pi_j\rho_i=0$ for any $i\not=j$.
\end{Lemma}

{\it Proof.} Suppose for any $i$, $\Pi_i$ can unambiguously
discriminate $\rho_i$, then we have $\tr(\Pi_j
\rho_i)=p_i\delta_{ij}$ for some $p_i>0$. Let
\begin{equation}\label{equ:decomposition}
\rho_i=\sum_{k=1}^{n_i}r_i^k \ket{\psi_i^k}\bra{\psi_i^k}
\end{equation}
for some $r_i^k>0$ be the spectrum decomposition of $\rho_i$, then
for any $i\not =j$
\begin{equation}
0=\tr(\Pi_j \rho_i)=\sum_{k=1}^{n_i}r_i^k \bra{\psi_i^k}\Pi_j
\ket{\psi_i^k}
\end{equation}
and $\bra{\psi_i^k}\Pi_j\ket{\psi_i^k}=0$ for $k=1,\dots,n_i$ from
the fact $r_i^k>0$. That implies $\Pi_j\ket{\psi_i^k}=0$ and so
$\Pi_j \rho_i=0$.
 \hfill $\blacksquare$

\vspace{1em}

It is well known that perfect pure state discrimination is
possible if and only if the states to be discriminated are
orthogonal to each other. In the case of mixed state, we have a
similar result as the following theorem.

\begin{Theorem} The quantum mixed states $\rho_1,\rho_2,
\dots,\rho_n$ can be
perfectly discriminated if and only if they are orthogonal, that
is, $\rho_i\rho_j=\delta_{ij}\rho_i^2$.
\end{Theorem}

{\it Proof.} If $\rho_i\rho_j=\delta_{ij}\rho_i^2$, then
$\fsp(\rho_i)\perp\fsp(\rho_j)$ for any $i\not=j$. We choose
$\Pi_i$ as the projector onto \fsp($\rho_i$). Obviously $\sum_i
\Pi_i = I_s$ and $\tr(\Pi_i \rho_j)=\delta_{ij}$, where $I_s$ is
the identity matrix in \fsp($\rho_1,\dots,\rho_n$). That indicates
$\Pi_1,\dots,\Pi_n$ can perfectly discriminate
$\rho_1,\dots,\rho_n$.

Conversely, if $\rho_1,\rho_2, \dots,\rho_n$ can be discriminated
perfectly, then there exist POVM elements $\Pi_1,\dots,\Pi_n$,
$\sum_k \Pi_k=I$, such that for any $i=1,\dots,n$, $\Pi_i$ can
perfectly (so unambiguously) discriminate $\rho_i$. From Lemma
\ref{lemma:1}, we have $\Pi_j\rho_i=0$ for any $i\not=j$. So
$\rho_i \rho_j=\rho_i(\sum_k \Pi_k) \rho_j=\delta_{ij}\rho_i^2$.
\hfill $\blacksquare$

\vspace{1em}

The above theorem gives us a sufficient and necessary condition
when mixed states can be discriminated perfectly. That is, they
must be orthogonal to each other. In the case when the states are
nonorthogonal, a strategy is, as in pure state situation,
unambiguous discrimination. While a set of pure states can be
unambiguously discriminated if and only if they are linearly
independent \cite{CH98}, the unambiguous discrimination between
mixed states has a stronger requirement, as the following theorem
indicates.

\begin{Theorem}
The quantum mixed states $\rho_1,\dots,\rho_n$ can be
unambiguously discriminated if and only if for any $i=1,\dots,n$,
$\fsp(S)\not=\fsp(S_i)$, where $S=\{\rho_1,\dots,\rho_n\}$ and
$S_i=S\backslash \{\rho_i\}$.
\end{Theorem}
{\it Proof.} ``$\Longrightarrow$". Suppose $\rho_1,\dots,\rho_n$
can be unambiguously discriminated, then there exist POVM elements
$\Pi_1,\dots,\Pi_n$ such that $\sum_i \Pi_i \leq I$ and $\tr(\Pi_i
\rho_j)=p_i\delta_{ij}$ for some $p_i>0$. Let $\ket{\psi_i^k},
k=1,\dots,n_i$, be the eigenvectors of $\rho_i$ with the
corresponding eigenvalues larger than 0. Then there exists $1\leq
h_i\leq n_i$ such that
$\bra{\psi_i^{h_i}}\Pi_i\ket{\psi_i^{h_i}}>0$ and from Lemma
\ref{lemma:1}, for any $1\leq j\leq n_i$,
$\bra{\psi_i^j}\Pi_k\ket{\psi_i^j}=0$ provided that $i\not = k$.

In what follows, we prove that for any $i=1,\dots, n$,
$\ket{\psi_i^{h_i}}$ cannot be written as a linear combination of
the states $\ket{\psi_k^j}$ for $k\not =i$ and $j=1,\dots,n_k$,
that will imply the result $\fsp(S_i)\not=\fsp(S)$. Suppose
$$\ket{\psi_i^{h_i}}=\sum_{k\not=i,j}a^i_{k,j}\ket{\psi_k^j}$$ for
some $a^i_{k,j}$, then
\begin{equation}
\Pi_i\ket{\psi_i^{h_i}}=\sum_{k\not=i,j}
a^i_{k,j}\Pi_i\ket{\psi_{k}^{j}}=0,
\end{equation}
which contradicts with
$\bra{\psi_i^{h_i}}\Pi_i\ket{\psi_i^{h_i}}>0$.

``$\Longleftarrow$". Suppose $\fsp(S)\not=\fsp(S_i)$, then
$\fsp(\rho_i)\not\subseteq \fsp(S_i)$. It follows that there
exists a state $\ket{\phi_i}$ such that $\ket{\phi_i}\not \perp
\fsp(\rho_i)$ but $\ket{\phi_i}\perp \fsp(S_i)$. That is
$\bra{\phi_i}\rho_i\ket{\phi_i}>0$ but
$\bra{\phi_i}\rho_k\ket{\phi_i}=0$ for any $k\not=i$. Let
$\Pi_i=q_i\ket{\phi_i}\bra{\phi_i}$, where $q_i$ is sufficient
small but positive such that $\sum_{i=1}^n \Pi_i\leq I$, we can
check easily that the POVM elements $\Pi_1,\dots,\Pi_n$ can
unambiguously discriminate $\rho_i$ with a positive probability
$p_i=q_i\bra{\phi_i}\rho_i\ket{\phi_i}>0$ for any $i=1,\dots,n$.
\hfill $\blacksquare$

\vspace{1em}

When $\rho_1,\dots,\rho_n$ are all pure states, the requirement of
them to be unambiguously distinguishable presented in the above
theorem is exactly that they should be linearly independent, just
as we all know. This is because if
$\rho_i=\ket{\psi_i}\bra{\psi_i}$ for some state $\ket{\psi_i}$
then $\fsp(S)\not=\fsp(S_i)$ for any $i$ if and only if
$\ket{\psi_1},\dots,\ket{\psi_n}$ are linearly independent.

In general, however, the requirement of $\rho_1,\dots,\rho_n$ to
be unambiguously distinguishable is more strict than just linear
independence. To see this, for any $i=1,\dots,n$, suppose
$\fsp(S)\not=\fsp(S_i)$, we show $\rho_i$ cannot be written as a
linear combination of $\rho_j$, where $j\not=i$. In fact, if
$\rho_i=\sum_{j\not=i} a_j^i \rho_j$ for some $a_j^i$, let
$\ket{\phi_i}$ be a state orthogonal to $\fsp(S_i)$ but not
orthogonal to $\fsp(\rho_i)$, then
$$0<\bra{\phi_i}\rho_i\ket{\phi_i}=\sum_{j\not=i} a_i^j \bra{\phi_i}\rho_j\ket{\phi_i}
=0.
$$
This contradiction indicates that $\rho_1,\dots,\rho_n$ are
linearly independent. The converse, however, does not necessarily
hold. That is, the linear independence of $\rho_1,\dots,\rho_n$
cannot guarantee that $\fsp(S)\not=\fsp(S_i)$. To see this, let us
give a simple example. Suppose $\rho_1$ and $\rho_2$ are two
different density matrices with rank $m$ in an $m$-dimensional
Hilbert space. It is obvious that $\rho_1$ and $\rho_2$ are
linearly independent but
$\fsp(\rho_1)=\fsp(\rho_2)=\fsp(\rho_1,\rho_2)$. So in general the
linear independence of certain mixed states cannot ensure the
existence of a POVM to unambiguously discriminate between them.

We now turn to consider the problem of unambiguously
discriminating between $n$ quantum mixed states with a prior
probabilities. The aim is to optimize the discrimination by
choosing appropriate measurements to maximize the success
probability, or equivalently, minimize the inconclusive
probability. For the general case of unambiguous discrimination
between $n$ pure states, the optimization problem can be reduced
to a semidefinite problem \cite{SZ02}, which has no analytic
solution. So the bound on the success (or inconclusive)
probability for any unambiguous discrimination process becomes
very important. A lot of works such as Ref. \cite{ZF01} and Ref.
\cite{FZ02} dedicate to this field. In the following, we derive a
lower bound on the inconclusive probability of unambiguous
discrimination between $n$ mixed states using a method similar to
that in Ref. \cite{FZ02}.

\begin{Theorem}
Suppose a quantum system is prepared in one of the $n$ mixed
states $\rho_1,\dots,\rho_n$ with $a$ $prior$ probabilities
$\eta_1,\dots,\eta_n$. Then a lower bound on the inconclusive
probability $P_0$ of unambiguous discrimination between these
states is
$$P_0\geq \sqrt{\frac{n}{n-1}\sum_{i\not=j}\eta_i\eta_j F(\rho_i,\rho_j)^2}$$
where $F(\rho_i,\rho_j)$ is the fidelity of $\rho_i$ and $\rho_j$.
\end{Theorem}

{\it Proof.} For any POVM elements $\Pi_1,\dots,\Pi_n$, $\sum_i
\Pi_i\leq I$, which can unambiguously discriminate
$\rho_1,\dots,\rho_n$, we have $\tr(\Pi_i\rho_j)=p_i\delta_{ij}$
for $i,j=1,\dots,n$. Define $\Pi_0=I-\sum_{i=1}^n \Pi_i\geq 0$,
then $P_0=\sum_i \eta_i \tr(\Pi_0\rho_i)$. So

\begin{equation}\label{equ:P02}
P_0^2=\sum_i \eta_i^2
(\tr(\Pi_0\rho_i))^2+\sum_{i\not=j}\eta_i\eta_j
\tr(\Pi_0\rho_i)\tr(\Pi_0\rho_j).
\end{equation}
By Cauchy inequality, we have
\begin{equation}\label{equ:ca2}
\sum_i \eta_i^2 (\tr(\Pi_0\rho_i))^2\geq
\frac{1}{n-1}\sum_{i\not=j}\eta_i\eta_j
\tr(\Pi_0\rho_i)\tr(\Pi_0\rho_j).
\end{equation}
Substituting Eq.(\ref{equ:ca2}) into Eq.(\ref{equ:P02}) we have
\begin{equation}\label{equ:P021}
P_0^2\geq \frac{n}{n-1}\sum_{i\not=j}\eta_i\eta_j
\tr(\Pi_0\rho_i)\tr(\Pi_0\rho_j).
\end{equation}
Furthermore, using Cauchy inequality again, we have
\begin{equation}
\begin{array}{rl}
&\tr(\Pi_0 \rho_i)\tr(\Pi_0 \rho_j) \\
\\
&=\tr(U\sqrt{\rho_i}\sqrt{\Pi_0} \sqrt{\Pi_0}
\sqrt{\rho_i}U^\dagger)\tr(\sqrt{\rho_j}\sqrt{\Pi_0} \sqrt{\Pi_0}
\sqrt{\rho_j})\\
\\
&\geq
(\tr(U\sqrt{\rho_i}\Pi_0\sqrt{\rho_j}))^2\\
\\
&=(\tr(U\sqrt{\rho_i}(I-\sum_{k=1}^n \Pi_k) \sqrt{\rho_j}))^2
\end{array}
\end{equation}
for any unitary matrix $U$. From Lemma \ref{lemma:1}, we have
$\sqrt{\rho_i}\Pi_k\sqrt{\rho_j}=0$ for any $i\not=j$ and
$k=1,\dots,n$. Notice also that
$$F(\rho_i,\rho_j)=\max_{U} \tr(U\sqrt{\rho_i}\sqrt{\rho_j})$$
where the maximum is taken over all unitary matrix $U$. It follows
that for any $i\not=j$,
\begin{equation}\label{equ:sqrt}
\tr(\Pi_0 \rho_i)\tr(\Pi_0 \rho_j)\geq F(\rho_i,\rho_j)^2
\end{equation}
Taking Eq.(\ref{equ:sqrt}) back into Eq.(\ref{equ:P021}) we derive
the lower bound on $P_0$ as
\begin{equation}\label{equ:bound}
P_0\geq \sqrt{\displaystyle\frac{n}{n-1}\sum_{i\not=j}\eta_i\eta_j
F(\rho_i,\rho_j)^2}.
\end{equation}
That completes the proof of this theorem. \hfill $\blacksquare$

\vspace{1em}

When $n=2$, the lower bound we presented above reduces to $P_0\geq
2\sqrt{\eta_1\eta_2}F(\rho_1,\rho_2)$, which partially coincides
with the bound given in \cite{RS03}. On the other hand, when
$\rho_1,\dots,\rho_n$ are all pure states, then the lower bound
reduces to the one derived in Ref. \cite{FZ02}.

What we would like to point out here is that from the proof of the
above theorem, we can derive a series of lower bounds on the
inconclusive probability. In fact, if let
$$A_k=\sum_i \eta_i^{2k} (\tr(\Pi_0\rho_i))^{2k}$$ and
$$B_k=\sum_{i\not=j} \eta_i^k\eta_j^k (\tr(\Pi_0\rho_i))^k(\tr(\Pi_0\rho_j))^k$$
then by Cauchy inequality, we have $A_{k}\geq B_k/(n-1)$. Using
these notations, the key steps Eq.(\ref{equ:P02})-(\ref{equ:P021})
in the proof of the above theorem can be reexpressed as
\begin{equation}\label{equ:P022}
P_0^2=A_1+B_1\geq \frac{n}{n-1}B_1
\end{equation}
which implies the lower bound
\begin{equation}\label{equ:bound1}
P_0\geq P_0^{(1)}\doteq\sqrt{\frac{n}{n-1}C_1}
\end{equation}
as in Eq.(\ref{equ:bound}), where $C_k$ is defined by
$$C_k=\sum_{i\not=j} \eta_i^k\eta_j^k F(\rho_i,\rho_j)^{2k}$$

Now, if we notice the fact that $A_k^2=A_{2k}+B_{2k}$, then we can
first consider the term $A_1$ and derive that
$A_1=\sqrt{A_2+B_2}$, so we can rewrite Eq.(\ref{equ:P022}) as
$$P_0^2=\sqrt{A_2+B_2}+B_1\geq \sqrt{\frac{n}{n-1}B_2}+B_1.$$
which implies another lower bound
\begin{equation}\label{equ:bound2}
P_0\geq P_0^{(2)}\doteq\sqrt{C_1+\sqrt{\frac{n}{n-1}C_2}}.
\end{equation}
One can easily prove that the bound presented in
Eq.(\ref{equ:bound2}) is better than that in Eq.(\ref{equ:bound1})
by Cauchy inequality.

Similarly, we can derive a series of lower bounds on the
inconclusive probability of unambiguous discrimination between $n$
mixed states as follows
\begin{equation}\label{equ:boundk}
P_0\geq P_0^{(k)}\doteq
\sqrt{C_1+\sqrt{\dots+\sqrt{\frac{n}{n-1}C_k}}}.
\end{equation}
We can also prove that $P_0^{(1)}\leq P_0^{(2)}\leq \dots$, that
means when $k$ increases, the lower bounds become better and
better in the sense that they are closer and closer to the real
optimal inconclusive probability. On the other hand, since the
increasing sequence $\{P_0^{(k)}, k=1,2,\dots\}$ has an upper
bound 1, they definitely converge at a limit $P_0^{(\infty)}$,
which is the best lower bound we can derive using this method.

To summarize, we prove that any state chosen from a mixed state
set can be perfectly discriminated if and only if the set are
orthogonal. For the case of nonorthogonal mixed state set, the
sufficient and necessary condition of when states from it can be
unambiguously discriminated is that any state in the set has the
support space not totally included in the supports of the others.
We consider also the problem of discriminating unambiguously of
$n$ mixed states with \textit{a prior} probabilities and present a
series of lower bounds on the inconclusive probability.

\smallskip
{\bf Acknowledgement:} The authors thank S. Y. Zhang for useful
discussion in this field. In pure state case, he pointed out a
result similar to the series of lower bounds presented in this
paper. This work was supported by National Foundation of Natural
Sciences of China (Grant No: 60273003), National Key Project for
Fundamental Research of China (Grant No: 1998030905).


\begin{thebibliography}{9999}

\bibitem{IV87} I.D. Ivanovic, Phys. Lett. A 123, 257 (1987).

\bibitem{DI88} D. Dieks, Phys. Lett. A 126, 303 (1988).

\bibitem{PE88} A. Peres, Phys. Lett. A 128, 19 (1988).

\bibitem{JS95} G. Jaeger, A. Shimony, Phys. Lett. A 197, 83(1995).

\bibitem{CH98} A. Chefles, Phys. Lett. A 239, 339(1998).

\bibitem{SZ02} X. M. Sun, S. Y. Zhang, Y. Feng, M. S. Ying, Phys. Rev. A 65, 044306(2002).

\bibitem{EL03} Y. C. Eldar, IEEE Trans. Infom. Theory, Vol 49, No. 2,
2003.

\bibitem{ZF01} S. Y. Zhang, Y. Feng, X. M. Sun, M. S. Ying, Phys. Rev. A 64, 062103(2001).

\bibitem{FZ02} Y. Feng, S. Y. Zhang, R. Y. Duan, and M. S. Ying, Phys. Rev. A 66, 062313 (2002).

\bibitem{SB02} Y. Sun, J.A. Bergou, and M. Hillery, Phys. Rev. A 66, 032315
(2002).

\bibitem{RS03} T. Rudolph, R. W. Spekkens, P. S. Turner, Phys. Rev. A 68, 010301(R)(2003).

\bibitem{RL03} P. Raynal, N. Lutkenhaus, S. J. van Enk, Phys. Rev. A 68,
022308 (2003).

\bibitem{FJ03} J. Fiurasek and M. Jezek, Phys. Rev. A 67, 012321 (2003).

\bibitem{EL031} Y. C. Eldar, Phys. Rev. A 67, 042309 (2003).

\end{thebibliography}
\end{document}